# A Wearable Data Collection System for Studying Micro-Level E-Scooter Behavior in Naturalistic Road Environment

**Avinash Prabu, Dan Shen, Renran Tian, Stanley Chien, Lingxi Li, Yaobin Chen**
**Transportation & Autonomous Systems Institute (TASI), IUPUI**

*Corresponding Author: Renran Tian*
*ET 301, 799 W. Michigan St.*
*Indianapolis, Indiana, 46202, USA*
*Phone: 1 317 278-8717*
*Fax: 1 317 274-4567*
*Corresponding author's e-mail: rtian@iupui.edu*

**Rini Sherony**
**Collaborative Safety Research Center (CSRC), Toyota Motor North America (TMNA)**

1555 Woodridge Avenue, Ann Arbor, MI 48105, USA
Phone: (1) 734-995-7000

**ABSTRACT**: As one of the most popular micro-mobility options, e-scooters are spreading in hundreds of big cities and college towns in the US and worldwide. In the meantime, e-scooters are also posing new challenges to traffic safety. In general, e-scooters are suggested to be ridden in bike lanes/sidewalks or share the road with cars at the maximum speed of about 15-20 mph, which is more flexible and much faster than the pedestrains and bicyclists. These features make e-scooters challenging for human drivers, pedestrians, vehicle active safety modules, and self-driving modules to see and interact. To study this new mobility option and address e-scooter riders' and other road users' safety concerns, this paper proposes a wearable data collection system for investigating the micro-level e-Scooter motion behavior in a Naturalistic road environment. An e-Scooter-based data acquisition system has been developed by integrating LiDAR, cameras, and GPS using the robot operating system (ROS). Software frameworks are developed to support hardware interfaces, sensor operation, sensor synchronization, and data saving. The integrated system can collect data continuously for hours, meeting all the requirements including calibration accuracy and capability of collecting the vehicle and e-Scooter encountering data.

## 1. Introduction

As one of the most popular micro-mobility options, e-scooters are spreading in hundreds of big cities and college towns in the US and worldwide. According to multiple nation-wide surveys, more than 70% of respondents across US cities report having positive attitudes and pleasant riding experiences towards these convenient, fun, and cost-efficient transportation tools. Companies such as Bird, Lime, Uber, and Lyft are increasing their fleets of rental e-scooters quickly.

In the meantime, e-scooters are also posing new challenges to traffic safety. In general, e-scooters are suggested to be ridden in bike lanes or share the road with cars. However, people may also ride e-scooters on sidewalks. In the former conditions, e-scooter riders face the risks of crashing with vehicles or bicyclists; while in the latter one, the e-scooters put more risks on pedestrians and themselves while performing avoidance maneuvers. Currently, there are no specified regulations established for this new type of mobility tool. Rental e-scooters can usually run at the maximum speed of about 15-20 mph. Comparing with other common road users, e-scooters are smaller than bicyclists, slower than cars, faster than pedestrians, and much more agile than all of them. Together, these features make e-scooters challenging for human drivers, pedestrians, vehicle active safety modules, and self-driving modules to see and interact. Thus, e-Scooters pose new challenges to vehicles' active safety and automated-driving systems, especially considering that they are much slower than cars, small and less visible to vehicle sensing systems, and can quickly change their movements. Common accidents with e-Scooters include:

- Pedestrians tripping over parked scooters
- Automobiles or trucks hitting riders
- Riders striking a pedestrian on the sidewalk
- Riders crashing when performing avoidance maneuvers
- e-Scooter defects/malfunctions
- Road hazards
- Users riding recklessly, while impaired, or intoxicated.

The number of crashes associated with e-scooters tend to increase quickly, highlighting the importance of carefully studying this new mobility option and addressing e-scooter riders' and other road users' safety concerns. Such research requires a large amount of behavior and interaction data of e-scooters used by the public in the naturalistic road environment, especially the micro-level movement and interaction data. There are three ways to conduct such data collection. The first approach is to use a car-based data collection system equipped with cameras and other sensors to record the behavior of e-scooters via naturalistic driving. This method can be inefficient to encounter a large number of e-scooters, especially in less busy road locations or time and days. The second approach uses road-side cameras or other sensors to collect accurate data but may miss details visible only from the car's view angles or shorter distances. More importantly, road-side cameras can only be installed in limited predetermined road locations and cannot capture different behaviors of e-scooters from a broader range of circumstances.

The last approach is to conduct e-scooter-based data collection. Comparing with the previous two methods, this approach has several advantages: (1) efficiency in encountering a large number of cars; (2) capability to collect data in all road locations; (3) direct measurements of the e-scooters' micro-level behavior with details; (4) accurate recording of surrounding objects, and (5) the possibility to record the riders' demographic backgrounds via controlled human subject experiments to better understand individual differences. Considering the small size, unique shape, and highly dynamic moving patterns of e-scooters, there is no existing sensing solution or data acquisition system to support this type of experiment. Hence, an e-scooter-based data collection system needs to be developed that can:

- Collect accurate motion dynamics of the e-scooter.
- Fit into different current and future e-scooter designs.
- Collect movement data of surrounding road users like cars and pedestrians to reconstruct the interactions and detect potential conflicts.

With these requirements, we select a 64-beam 3D LiDAR, HD cameras, an RTK (Real-Time Kinematic) GPS as the primary sensors to collect the moving patterns of the ego-e-scooter as well as surrounding vehicles and pedestrians. The RTK GPS can generate the e-scooter trajectory with an accuracy level of 2-10 cm. In this approach, satellites' signals are supported by base towers to localize the unit, while accumulated local IMU inputs can compensate for temporary GPS loss. LiDAR and camera data are fused to reconstruct 3D surrounding views, taking advantages of both the higher resolution of cameras and accurate 3D distance measurements from the LiDAR. Fusing these four sensors allows the data collection system to reconstruct the global trajectories of all moving objects in the view, record their detailed movements and state changes, and further calculate distances, velocities, directions, and time-to-collision (TTCs) among them.

All the sensors are integrated and synchronized into a wearable data collection system to fit all possible e-scooter designs. Considering the small sizes of e-scooters, it is challenging to mount all the components on the e-scooter, which may affect its maneuvers and be prone to the potential damages caused by falling and crashes. Also, e-scooter design evolves quickly, and it is not easy to design a universal data collection system to fit all of them. Thus, a wearable design is a better choice. There is no current wearable system that can meet this study's needs, so we developed an original one.

With its unique advantages as a mobility option, the increased popularity means there will be more and more e-Scooters on the road. The number of crashes and accidents associated with them also tends to increase quickly, which presents a new challenge to road safety for automated vehicles and other vulnerable road users. It is important to study this new mobility option carefully and address safety concerns of e-Scooter riders and other road users. To our best knowledge, there has not been systematic research done to focus on the micro-level moving behavior and crash scenarios related to e-Scooters with public use in an open road environment. Thus, we propose to conduct a research study investigating the micro-level behavior of e-Scooters in a naturalistic road environment.

## 2. Related Studies

Previous studies suggest that there are a few wearable data collection system in research and in the market right now, but there hasn't been a system that is designed specifically to cater to the needs of an e-Scooter rider. One of the closest researches in this area is by David Blankeau et al. [4], which developed a low-cost 2D LiDAR technology for data collection from bicycles. A high amount of processing power is needed to extract high speed motion information and thus does not suit our research problem. Another system that is very close to the proposed system in the Google Street View Trekker [5]. The system is designed to capture 360 degree camera data and 3D LiDAR data. The main focus of the system is to capture the topology for Google street view. This system also does not solve our research needs because of it's functionality and the height at which the sensors are mounted with respect to the user. A more complex system with a short range LiDAR in addition to a Laser range finder and a ZED stereo camera for autonomous three-wheeled scooters is proposed by Liu et al [6]. In this system the sensors are mounted on the scooter itself, which makes it unusable for our goals. Though a vehicle-based data collection system has been well documented in researches [7] – [12], a similar system cannot be used for e-Scooters due to the constraint on size and weight.

Similar architechtures that use cameras with an embedded system running Linux have been researched in [13], [14], and [15]. A large amount of research has been documented in the field of LiDAR camera systems in various applications such as forest data collection [16] and coastal mapping [17]. The other key point of LiDAR-camera-based perception system is the sensor fusion/calibration. The calibration process can synchronize the two sensor inputs through the coordinate transformation and data point mapping. Mnay researchers have done the great develoments on the LiDAR-camera fusion algorithms. An interactive LiDAR to camera calibration toolbox to estimate the intrinsic and extrinsic transforms has been introduced in [18]. The authors in [19] address the common, yet challenging, LiDAR-camera semantic fusion problems, which are seldom approached in a wholly probabilistic manner. A novel open-source tool for extrinsic calibration of radar, camera, and LiDAR has been presented in [20] with facilitating joint extrinsic calibration of all three sensing modalities for multiple measurements. [21] presents a pipeline for extrinsic calibration of a ZED stereo camera with a Velodyne Puck LiDAR. This pipeline uses a novel 3D marker whose edges can be robustly detected both in the image and 3D point cloud. In another recent study towards extrinsic calibration of LiDAR-Camera fusion, an algorithm has been introduced in [22] to estimate the similarity transformation between laser points and pixels using a checkerboard.

## 3. e-Scooter-based Data Acquisition System Development

The main purpose of the data acquisition system is to record real-time data of vehicles on the road. This section focuses on development of an e-Scooter based data acquisition system. Three types of sensors have been used in the data acquisition system: cameras for video data, LiDAR for distance and IMU data, and GPS for latitude, longitude, and altitude data. The embedded hardware system, for facilitating data recording, runs on Ubuntu/ROS platform. The sensors' data are stored in files with ROS bag format, which can be processed into raw sensor messages for further analysis and processing.

3.1. Sensor Selection

There are several constraints while choosing sensors and a data collection device to be mounted on an e-Scooter. Primary considerations are weight and duration of operation of the system. Since the system is designed to collect data for the surrounding objects, the final system consists of three USB cameras, a 3D LiDAR with integrated IMU, a RTK GPS unit, and a NVIDIA Jetson Tx2 development computer board for data collection.

3.1.1. LiDAR Selection

The main features considered while choosing a LiDAR were 2D vs. 3D, resolution, frame rate, sensor speed, number of distance points, hardware, accuracy, and cost.

2D LiDAR only captures the object in the plane of view of the sensor. Its incapability to capture data above and below the LiDAR plane led to choosing a 3D LiDAR. Comparing the currently available 3D LiDARs in the market, the Ouster OS-1 64 beam medium-range LiDAR was chosen. The choice was based on functionality to cost terms. The 64-beam variant was selected over the 32-beam variant due to the high number of data points that could be gleaned over the horizontal field of view of the LiDAR compared to the 32 beam variant. Another critical factor for choosing the OS-1 was, it comes with a sensor interface that allows transmitting data over Ethernet using UDP. The OS-1 can generate 1,310,720 points at a frequency of 10Hz, and has a horizontal field of 360 degrees, and a 45 degrees vertical field of view. The LiDAR has a maximum distance range of 120m with an accuracy of +/- 5cm for Lambertian targets. The Ouster OS-1 64 beam LiDAR is shown in Fig. 1.

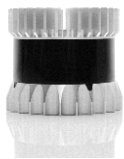

Fig. 1 Ouster OS-1 64 beam 3D LIDAR.

3.1.2. Camera Selection

The main features considered while selecting a camera were resolution, frames per second, shutter type, connector type, color coding, external hardware synchronization, data transfer speed, compatibility with ROS operating system, and cost. For portability reasons, the weight and dimensions of the camera were also important factors. The compatibility on the software side was also a consideration in camera selection. The two cameras in consideration were Sony IMX291 and Logitech c920. After several experiments, it was found that the Sony IMX291 was not capable of dynamic memory allocation, which causes a problem while having multiple cameras on the system. On the other hand, the Logitech c920 is easily programmable and is compatible with running multiple cameras with dynamic memory allocation. With all these considerations in mind, we selected the Logitech c920 webcam capable of delivering 1080p @30fps with h.264 or MJPEG format over USB3.0 (Fig. 2). The camera came inbuilt with a cs-mount lens and a horizontal field of view of 78 degrees, and a diagonal field of 90 degrees

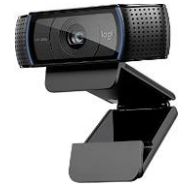

Fig. 2 Logitech c920

3.1.3. GPS Selection

GPS unit can record the e-Scooter-based data collection systems' position and motion profile. The overall structure of RTK GPS can be shown in Fig. 3. The system has three key components: 1. Emlid Reach M+ RTK GNSS Module as a rover, 2. Tallysman Global Navigation Satellite System (GNSS) Antenna, and 3. Indiana Department of Transportation coordinates a network of continuously operating reference stations (INDOT-INCORS) System as a station. In general, the base station receives a signal from GPS satellites, creates a correction factor, and then sends the correction signal via WIFI to the rover. The INDOT-INCORS system acts as the base station for the whole RTK GPS. Through the INDOT-INCROS System, each base station site provides Global Navigation Satellite System (GNSS - GPS and GLONASS) carrier phase and code range measurements to support 3-dimensional positioning activities throughout the state of Indiana. The rover is the Reach M+ RTK GNSS Module with 5GB internal storage, which can provide precise navigation (centimeter accuracy) and autonomous vehicle mapping. It has a built-in 9 DOF Inertial Measurement Unit (IMU) that can also obtain the car system's or the e-Scooter system's rotation dynamics. The Reach M+ can stream in National Marine Electronics Association (NMEA) or binary format to the device over Universal asynchronous receiver-transmitter (UART), Bluetooth, or WIFI. Our system uses WIFI for the communications between the Base station and rover. The Tallysman GNSS antenna is a compact wideband antenna that provides accurate reception for all upper L- band GPS, GLONASS, Beidou, and Galileo signals (L1, G1, B1, B1 BOC, B12, E1) and associated augmentation signals (WAAS, EGNOS, and MSAS SBAS).

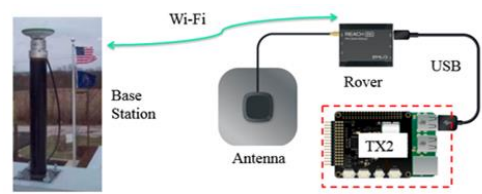

Fig. 3 Structure of RTK GPS

3.2 Data Collection Computer Selection

Since the onboard computer's volume, weight, and power consumption were primary considerations, we decided to go with an embedded system, preferably a single board computer, with appropriate storage capabilities to record data from the cameras and the LiDAR and store them efficiently. With these considerations in mind, we selected the NVIDIA Jetson Tx2, running an ARM-based Ubuntu 16.04 as the primary onboard computer (Fig. 4).

Although the Jetson Tx2 development kit comes with a relatively big motherboard, it fares higher in performance. The CPU contains a Dual-core NVIDIA Denver 2 in conjunction with an ARM Cortex A57 and 8GB of LPDDR4 memory. The CPU power and the memory size were the primary reasons for selecting the Jetson Tx2. The amount of memory allows for the ROS based LiDAR processes to use the RAM as a buffer space for incoming LiDAR data as the UDP packets is processed. The availability of a USB 3.0 port can be expanded using a USB hub for multiple camera operations. The presence of a gigabit Ethernet facilitates a full operation of the Ouster LiDAR. The SATA port was then used to connect to an external SSD to record the data coming in from the sensor.

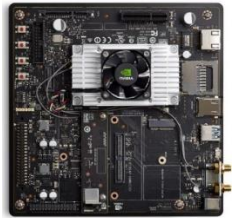

Fig. 4 Jetson Tx2 Developer Kit

3.3 Hardware System Structure

To run the data collection system wirelessly on-board, we used batteries to power the whole system. A standard laptop external charger was deemed a good option due to its pre-existing multiple voltage level selections and the higher than required energy capacity. The battery is capable of delivering 185W-h or 50,000mAh, with the ability to give out a maximum of 130Watts.The power bank could supply three voltages, 20V, 12 V, and 5V, with the latter being provided as a USB socket. The battery is 2.7lb, the system could go from 0 to full charge in under 6 hours. To provide the required current and voltage to the LiDAR, a DC-DC converter was used to provide the appropriate voltage. The Jetson Tx2 was directly connected to the battery's 20V port. The LiDAR's sensor interface was connected to the Jetson Tx2 through an Ethernet cable. The three cameras were connected to the Tx2's USB port through a Startech USB hub. The GPS module was also connected to the USB hub and shared the same power as the cameras.

A Samsung SSD 860 evo 1TB SATA was connected to the Jetson Tx2 through a SATA cable for data saving. SATA drives maxes at 600 MB per second when compared to the 400 MB per second for USB. This helps to avoid buffering of data coming in from the sensors. The overall harware structure is shown in Fig. 5

Since the requirement was a wearable DAS for the e-Scooter system, only one direction (180-200 degrees) was the maximum field of view. The cameras have to be carefully placed to cover the maximum available view around the rider.

The system can be worn either in the front or the back of the rider (Fig. 6 and Fig. 7).

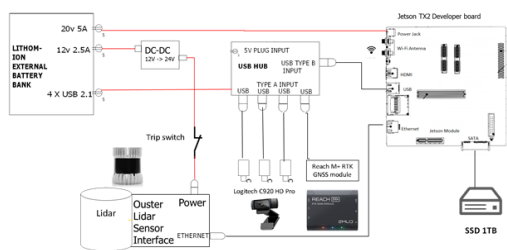

Fig. 5 Hardware System Structure

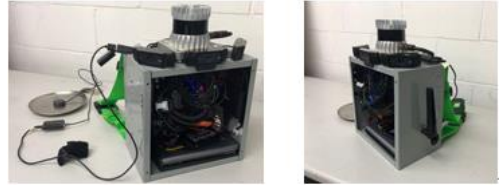

Fig. 6 Data Acquisition System

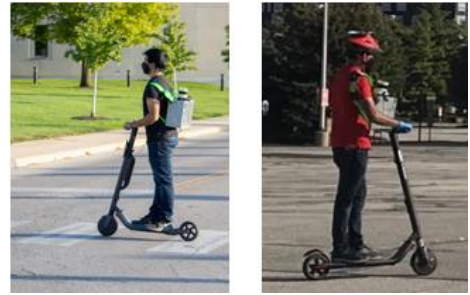

Fig. 7 System Orientation

3.4 System Design – Software

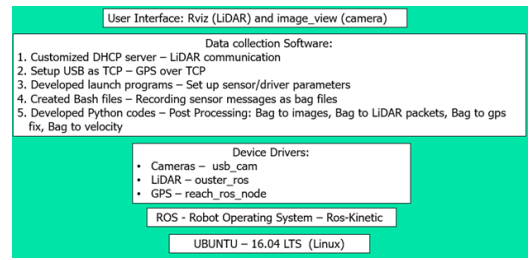

Fig. 8 Software Structure

Fig. 8 illustrates the software structure of the system. ROS-kinetic-1.12.14 was implemented on Ubuntu. ROS provides many advantages when it comes to data collection. One of the main factors for choosing ROS is the flexibility to control, record, and process multiple sensors simultaneously. Each sensor requires a device driver to capture and publish sensor message topics to be recorded. All the sensors that are selected have drivers made available by the manufacturer or third-party developers.

The LiDAR communicates with Jetson Tx2 using Ethernet. The LiDAR data is available through UDP, which requires the LiDAR to have a static IP address. This requirement is achieved by running a local DHCP server, which was primarily done with bash scripts and by including the command to run the script in the *bashrc* file in Ubuntu at the start of every reboot. The DHCP server setup primarily involved flushing the IP addresses on the Ethernet interface, followed by assigning a static IP address to the interface. Then it would be followed by the switching on the said interface, running the dnsmasq program (which started a DHCP server), and allowing devices physically connected to the Ethernet jack to get assigned IP addresses and establish the connection to transfer the LiDAR data. Once the connection is established, the device driver needs to be initiated. Ouster provides a maintained package for ROS variants, Kinetic and upwards, which is used to initiate and record the sensor messages.

The cameras connect to the Jetson Tx2 via USB 3.0. The *usb_cam* package of ROS is used to publish ROS image topics. The *usb_cam_node* interfaces with standard USB cameras using

*libusb_cam* and publishes images as *sensor_msgs: Image* (image messages in ROS format). *Libusb_cam* is a C library that provides generic access to USB devices. The *image_transport* library is used to allow compressed image transport. An image transport is a layer in ROS that supports transporting images in low bandwidth compressed formats like JPEG or PNG. To enable the operation, a launch file was created to initiate all the cameras. The launch file initiates the *usb_cam* driver and specifies the parameters that require to be set.

The GPS communicates with the Jetson Tx2 by Ethernet over USB configuration. Like the LiDAR, a static IP address is assigned to the GPS to establish communication with a computer, thereby assisting the *reach_ros_node* driver in publishing sensor messages for recording. The *reach_ros_node* uses a python script to run the GPS driver. The base station connection is achieved through the Reach View Web App, provided by the GPS manufacturer.

All the sensor messages are recorded as ROS bag files and then processed later to obtain raw sensor data. The operation of all the sensors at each step are integrated in bash scripts for ease of operation. The first bash script establishes connection with the sensors and allocates memory for optimal operation. The second bash script initiates all the sensors drivers and the third bash script records all sensor messages as bag files.

3.5 System Features

- Support continuous operation for 5 hours
- Total field of view: 200 degrees
- Frame rate: 10 Hz
- GPS accuracy: 2 Cm
- Portable and can be worn by the rider in front and back
- Weight ≈ 9 Kg

**4. System Calibration**

This section mainly describes the system calibration process and the calibration methods for the wearable data collection systems, the calibration results and accuracy, and the final data outputs from the system calibration systems.

4.1. Calibration Process

In general, the system calibration process for a 3D LiDAR and camera can be divided into two steps.
1. Find the intrinsic matrix, which purely relies on the camera itself.
2. Calculates the extrinsic matrix to build the connections between pixels in the camera image and the LiDAR 3D point cloud data.

The process of calibration involves primarily calculating three matrices of importance to achieve fusion. These three matrices are namely the "intrinsic matrix or the camera matrix," the "distortion parameters," and the "extrinsic transformation matrix." The camera matrix contains data representing camera parameters, such as focus, shear, and image width and height. The distortion coefficients represent the corrections due to lens distortion, such as for the fish-eye lens. These parameters are assumed to be zero for low field-of-view (FOV) lenses. In this project, we do not use the fish-eye lens and assume the distortion of the images is very limited (based on the manufacturer's manual). Therefore, in this application, we only consider two components of intrinsic camera matrix and extrinsic matrix in the system calibration and fusion, which is depicted in Figure 9. In the figure, from left to right, the 3D LiDAR point cloud data can be converted to 3D camera coordinates in the image plane using the extrinsic matrix. The intrinsic matrix will then be applied to the image plane for mapping the 3D LiDAR data to the 2D pixel coordinates in images. The sensor fusion process can be shown in the following intrinsic-extrinsic matrix equation, which governs LiDAR to camera coordinate transformation in the homogeneous representation.

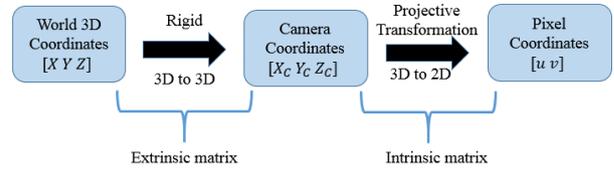

Fig 9. Process of system calibration [1].

The principle of LiDAR and camera calibration can be shown in equation 1.

$$\begin{bmatrix} u \\ v \\ 1 \end{bmatrix} = \begin{bmatrix} f_x & 0 & ox & 0 \\ 0 & f_y & oy & 0 \\ 0 & 0 & 0 & 1 \end{bmatrix} * \begin{bmatrix} r_{11} & r_{12} & r_{13} & t_x \\ r_{21} & r_{22} & r_{23} & t_y \\ r_{31} & r_{32} & r_{33} & t_z \\ 0 & 0 & 0 & 1 \end{bmatrix} * \begin{bmatrix} X \\ Y \\ Z \\ 1 \end{bmatrix} \quad (1)$$

where u and v are the pixel coordinates in the image plane, and the X, Y, and Z are the distance data from LiDAR measurements. The first matrix is a 3×4 camera intrinsic matrix, and the second one is a 4×4 extrinsic matrix. The detailed explanation of each component in the formula will be discussed in the following Sections.

4.2. Calculating the Camera Intrinsic Matrix

The camera intrinsic matrix projects 3D points given in the camera coordinate system to the 2D pixel coordinates [2] through a projection transformation, i.e.

$$\mathrm{p} = KP_c = \begin{bmatrix} f_x & 0 & ox \\ 0 & f_y & oy \\ 0 & 0 & 0 \end{bmatrix} P_c \quad (2)$$

where p is the 2D pixel coordinate, Pc is the 3D world coordinate in the camera coordinate system, and K matrix is the camera intrinsic matrix, which is a 3×3 matrix.

The matrix K is composed of the focal lengths $f_x$ and $f_y$, which are expressed in pixel units, and the principal point $(o_x, o_y)$, usually means the image center. The camera intrinsic matrix does not depend on the scene viewed. Thus, once calculated, it can be used as long as the focal length is fixed. However, if the image from the camera is scaled by a factor, all the parameters need to be scaled by the same factor.

The intrinsic camera matrix and the distortion parameters were calculated using MATLAB. The method needs the pictures of a checkerboard. We used a 9×6 checkerboard with each square having a 25mm×25mm dimension. For collecting the images, the checkerboard was placed in front of the camera at various angles and positions to cover the whole view. For the webcam, Gstreamer was used to capture the images using the Gstreamer command line. Gstreamer is a widely used open-source pipeline-based multimedia framework. The command makes sure that the new images would not be overwritten on top of the old images. Once 15-20 images of the checkerboard were collected, we can find the camera intrinsic matrix using the MATLAB Camera Calibrator App. As shown in Figure 10, all images were loaded and cycled in the App, and then the intrinsic matrix was generated, as well as the distortion parameters.

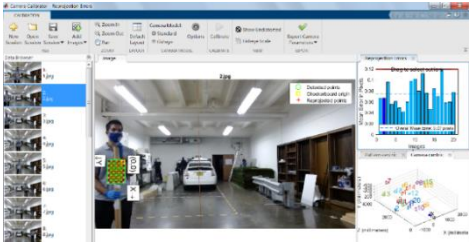

Fig 10. Intrinsic matrix calibration using MATLAB Camera Calibrator App.

### 4.3. Caculating the Extrinsic Matrix

The homogeneous transformation is encoded by the extrinsic matrix $\begin{bmatrix} R_{3\times 3} & T_{3\times 1} \\ \vec{0}^T & 1 \end{bmatrix}$, which includes two parts rotation R and translation T. These two extrinsic parameters represent the change of basis from a 3D (LiDAR) world coordinate system w to the 3D camera coordinate system c. Thus, given the representation of a point $p_w$ in the (LiDAR) world coordinate system, we can find the representation of a point $p_c$ in the camera coordinate system as follows:

$$P_c = \begin{bmatrix} R_{3\times 3} & T_{3\times 1} \\ \vec{0}^T & 1 \end{bmatrix} P_w = \begin{bmatrix} r_{11} & r_{12} & r_{13} & t_x \\ r_{21} & r_{22} & r_{23} & t_y \\ r_{31} & r_{32} & r_{33} & t_z \\ 0 & 0 & 0 & 1 \end{bmatrix} P_w \quad (3)$$

This homogeneous transformation is composed of a 3×3 rotation matrix R and a 3×1 translation vector T in the equation.

The process of obtaining the extrinsic matrix involved calculating an initial extrinsic matrix using a well-established algorithm known as the Levenberg-Marquardt algorithm. It is also known as the damped least-squares method for solving AX=B matrix equations, where X is the extrinsic matrix. The algorithm was implemented using the OpenCV SolvePnP function. After getting this initial matrix, a Python script was used to fine-tune the extrinsic matrix slightly.

*Initial Calibration*

The inputs of OpenCV SolvePnP function include the intrinsic matrix, the distortion parameters, and at least six pairs of image pixel values with corresponding LiDAR point cloud coordinate values. To collect the pairs in both the camera images and LiDAR point clouds for calculating the initial extrinsic matrix, three poster boards were set up at different distances from the cameras and LiDAR. The relative positions and angles of the cameras and the LiDAR are solidly fixed before the calibration process.

The data collection system was then used to record calibration data of these poster boards. This process was repeated for different cameras as the setup was kept roughly 3, 6, and 9 meters away from the LiDAR. The aim here was to obtain the point and pixel pairs of the LiDAR and camera data, respectively. The poster board was chosen primarily for its sharp vertices (four corners), which could be detected both in the LiDAR point cloud and the camera image. The poster boards were evenly spread out over the width of the image. It was observed that selecting points only from a specific region of the image could result in lousy fusion on other image areas, notably in width.

Once data were collected, these vertices' point and pixel values were used as inputs to the SolvePnP function. The pixel values were attained by opening the image file collected using an image app such as MS-Paint or Linux ImageMagick and finding out the pixel values of all the corners of the posters.

To find the corresponding 3D LiDAR points of these post board corners, a user interface via Jupyter Notebook was used to select them from the point cloud data. A Python script was utilized to convert the text files generated by the bag extraction script into a data frame containing X, Y, and Z values of the LiDAR data. These data frames were then plotted as a scatter plot using the Python Plotly library. The scatter plot then allowed for selecting a point (the poster edges) by simply hovering the cursor on the top of the point cloud as shown in Figure 11, and a label would display the XYZ values of that point [3].

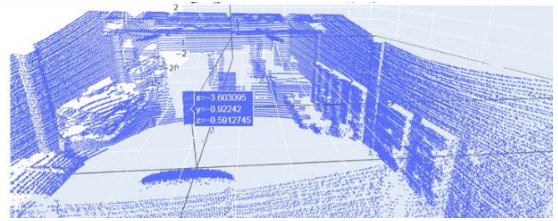

Fig. 11. User interface for selecting LiDAR points.

Three or six poster boards kept at different distances generated a total of 12 to 24 pixel-point pairs, which were then supplied to the SolvePnP function along with the camera intrinsic matrix and the distortion coefficients. The function would then return two vectors, each of them of a size 3×1. These vectors are called the rotation vector and the translation vector for the extrinsic matrix.

*Fine-tuning of the Extrinsic Matrix*

Once an extrinsic matrix was obtained, it was then manually refined with a Python script's aid. In this fine-tuning process, four poster boards were placed at various distances, as shown in Figure 12 below. The LiDAR and camera data were then collected and preprocessed through the same process as outlined earlier.

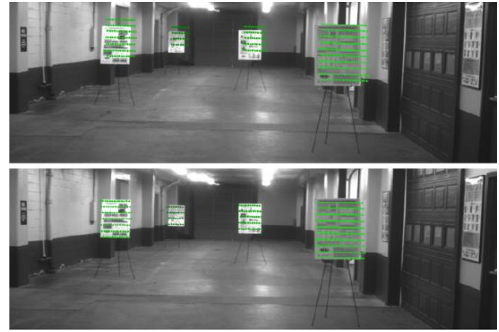

Fig. 12. Fusion effects before (Up) and after (Down) fine-tuning manually

Then, the collected point cloud data and images were loaded. The camera intrinsic matrix, the distortion parameters, the rotation matrix, and the translation vectors were input into the script. The script then performed a fusion calculation to project the LiDAR points onto the images (following Equation 1) and displayed the fused image [3].

### 4.4. Evaluation of the Calibration Results

The accuracy of the fusion results was checked by using the OpenCV. It allows to give a visual understanding of the accuracy of both matrices. The inputs to the function are the LiDAR points, the camera matrix, the distortion parameters, and the rotation and translational vectors. The function's outputs were the calculated 2D points containing u and v pixel values corresponding to the inputted LiDAR points.

The fusion was evaluated on the calibration data and some collected experimental data. Each generated 2D pixel value that fell inside the image's boundaries was displayed using a green dot on the image to represent the fusion overlay. Meanwhile, the True Positive Rate (TPR) metric was also calculated to evaluate the system calibration accuracy for the wearable data collection system within 5% error. For example, if the ground truth is 10m, the distances in 9.95 m to 10.05 m are considered correct mapping. Otherwise, they are wrong mappings. TPR can be calculated using the following formula:

$$True\ Positive\ Rate\ (TPR) = \frac{TP}{TP+TN} \quad (4)$$

where TP is the number of corrected mappings, and the TN is the number of wrong mappings. Therefore, the higher value of TPR, the better the fusion accuracy. Both the fusion results and the fusion accuracy will be discussed for each camera-LiDAR pair in the following subsections.

*LiDAR - Camera 1*

The final extrinsic matrix and the fused image were also obtained using the SolvePnP function in OpenCV, as shown in Figure 13. The True Positive Rate of checking the fusion accuracy of LiDAR-Camera 1 is shown with the ground truth of the green-colored poster board at 5.9 m, and the TRP=TP/(TP+TN)=100% within 5% error. The green dots mean all the projected points are in the boundary of the poster board.

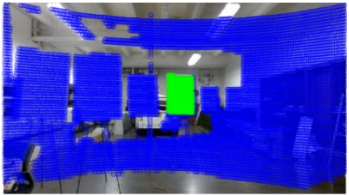

Fig. 13. Fusion accuracy of LiDAR-Camera 1.

*LiDAR - Camera 2*

The True Positive Rate of checking the fusion accuracy of LiDAR-Camera 2 is shown in Figure 14 with the ground truth of the green-colored poster board at 3.8 m, and the TRP= TP/(TP+TN)=100% within 5% error. The green dots mean all the projected points are in the boundary of the poster board.

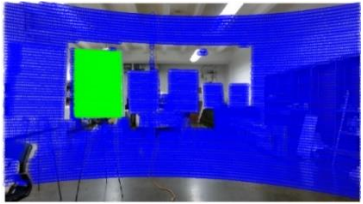

Fig. 14. Fusion accuracy of LiDAR-Camera 2.

*LiDAR - Camera 3*

The True Positive Rate of checking the fusion accuracy of LiDAR-Camera 3 is shown in Figure 15 with the ground truth of the green-colored poster board at 9.6 m, and the TRP=TP/(TP+TN)=100% within 5% error. The green dots mean all the projected points are in the boundary of the poster board.

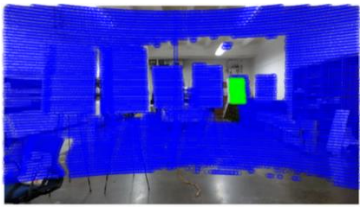

Fig. 15. Fusion accuracy of LiDAR-Camera 3.

## 5. Pilot Naturalistic Data

In this section, a pilot naturalistic study is described using the developed data acquisition systems. The wearable data collection system were used in the real road environment to collect the vehicle and e-Scooter encountering data.

The e-Scooter rider wearing the data acquisition system recorded naturalistic data of interactions with vehicles on the road. The duration of the recording is 47 minutes. The locations of the data collection were around downtown Indianapolis and near the IUPUI campus. The data collection started on September 26, 2020, at 3:23 PM. It had 36 minutes of back-facing and 11 minutes of front-facing data. The total size of data collected was 48.3 GB, and the total size of extracted and processed data was 90.5 GB. The travel route is given in Figure 16.

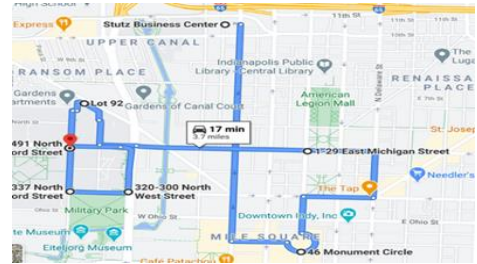

Fig. 16. Travel routes for wearable naturalistic data collection.

*Case 1 - The Front-facing In-parallel Case*

The first case of the e-Scooter-based naturalistic diving scenario was recorded on Indiana Avenue. As shown in Figure 17, two onboard cameras captured the target vehicle (Object 31 showing in the image) movements in front of the e-Scooter when the e-Scooter tried to stop slowly in the right lane. Initially, the vehicle was decelerating at the red light. Several seconds later, the vehicle was nearly stopped when the e-Scooter was decelerating behind the vehicle. When their distance was close enough, the e-Scooter adjusted its heading slightly and overtook the vehicle from the right side to avoid the crash. Since the most dangerous condition is when the e-Scooter was trying to regulate the heading angle when approaching the front vehicle, we consider this case as the in-parallel case.

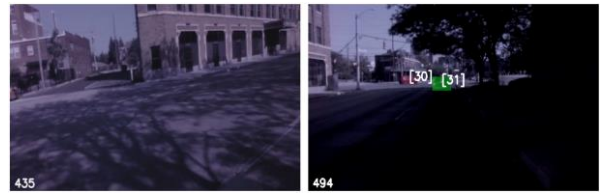

(a)    Front-left camera    (b) Front-middle camera

Fig. 17. Captured images of case 1 from the wearable data collection system.

*Case 2 - The Front-facing and In-parallel Case*

The second case of the e-Scooter-based naturalistic diving scenario has been recorded on St. Claire Street. As shown in Figure 18, the onboard two cameras can capture the target vehicle (Object 30 showing in the image) movements in front of the e-Scooter when the e-Scooter tried to stop slowly in case 2. Initially, the vehicle was decelerating when seeing the red light. Several seconds later, the vehicle was nearly stopped when the e-Scooter was decelerating behind the vehicle. When their distance was close enough, the e-Scooter adjusted its heading slightly and slowly approached the vehicle. Finally, the front vehicle started to accelerate and turned right when it is the green light. The most dangerous condition in this scenario is the moment that the e-Scooter was first decelerating and regulating the heading angle when approaching the front vehicle.

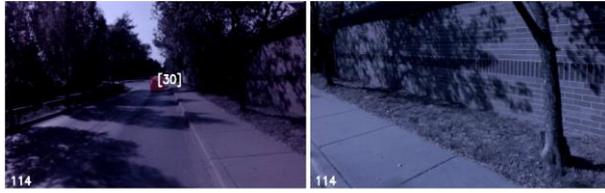

(a) Front-middle camera     (b) Front-right camera

Fig. 18. Captured images of Case 2 from the wearable data collection system.

## 6. Conclusion

This paper presents a wearable data collection system for studying the micro-level behavior of e-scooters in a naturalistic road environment. Facing the increased safety concerns related to this new type of micro-mobility device, the developed system can collect not only the moving patterns of e-scooters being ridden by different people but also the surrounding road users and the interactions among all of them. Through an appropriate process of data collection and sensor fusion towards ridding cases and interactions with crashes or near-crashes, researchers can better understand the risks and typical moving patterns of e-scooters in different road situations. This information can help develop vehicle active safety features, improve e-scooter designs, and create new traffic policies.

## References

(1) De Alvis, C., Shan, M., Worrall, S., and Nebot, E., 2019, May. Uncertainty Estimation for Projecting Lidar Points onto Camera Images for Moving Platforms. In 2019 International Conference on Robotics and Automation (ICRA) (pp. 6637-6643). IEEE.
(2) Wenyi Zhao, Rama Chellappa, P Jonathon Phillips, and Azriel Rosenfeld. Face recognition: A literature survey. Acm Computing Surveys (CSUR), 35(4):399–458, 2003.
(3) Betrabet, Siddhant Srinath (2021): Data Acquisition and Processing Pipeline for E-Scooter Tracking Using 3D LIDAR and Multi-Camera Setup. Purdue University Graduate School. Thesis. https://doi.org/10.25394/PGS.13342079.v1
(4) Blankenau, I., Zolotor, D., Choate, M., Jorns, A. et al., "Development of a Low-Cost LIDAR System for Bicycles," SAE Technical Paper 2018-01-1051, 2018, https://doi.org/10.4271/2018-01-1051.
(5) Cheung, Danny. "Mapping Stories with a New Street View Trekker." Google, 18 Dec. 2018, https://blog.google/products/maps/mapping-stories-new-street-view-trekker/.
(6) Liu, Kaikai, and Rajathswaroop Mulky. "Enabling Autonomous Navigation for Affordable Scooters." Sensors (Basel, Switzerland), MDPI, 5 June 2018, www.ncbi.nlm.nih.gov/pmc/articles/PMC6022038/.
(7) Pomerleau, Dean. \ALVINN: An Autonomous Land Vehicle in a Neural Network." NIPS (1988).
(8) A. Frome et al., "Large-scale privacy protection in Google Street View," 2009 IEEE 12th International Conference on Computer Vision, Kyoto, 2009, pp. 2373-2380, doi: 10.1109/ICCV.2009.5459413.
(9) "StreetSide: Dynamic Street-Level Imagery - Bing Maps." StreetSide: Dynamic Street-Level Imagery - Bing Maps, 12 Feb. 2020, www.microsoft.com/en-us/maps/streetside. doi: 10.1109/ICTIS.2017.8047822.
(10) A. Geiger, P. Lenz and R. Urtasun, "Are we ready for autonomous driving? The KITTI vision benchmark suite," 2012 IEEE Conference on Computer Vision and Pattern Recognition, Providence, RI, 2012, pp. 3354-3361. doi: 10.1109/CVPR.2012.6248074.
(11) Team, Waymo. \Waymo Open Dataset: Sharing Our Self-Driving Data for Research." Medium, Waymo, 21 Aug. 2019, medium.com/waymo/waymo-opendataset-6c6ac227ab1a.
(12) D. L. Rosenband, "Inside Waymo's self-driving car: My favorite transistors," 2017 Symposium on VLSI Circuits, Kyoto, 2017, pp. C20-C22.
(13) C. Yao-yu, L. Yong-lin, L. Ying and W. Shi-qin, "Design of image acquisition and storage system based on ARM and embedded," 2012 2nd International Conference on Consumer Electronics, Communications and Networks (CECNet), Yichang, 2012, pp. 981-984, doi: 10.1109/CECNet.2012.6202208.
(14) Y. Chai and J. Xu, "Design of Image Acquisition and Transmission System Based on STM32F407," 2018 2nd IEEE Advanced Information Management, Communicates, Electronic and Automation Control Conference (IMCEC), Xi'an, 2018, pp. 1085-1089, doi: 10.1109/IMCEC.2018.8469227.
(15) G. Sundari, T. Bernatin and P. Somani, "H. 264 encoder using Gstreamer," 2015 International Conference on Circuits, Power and Computing Technologies [ICCPCT-2015], Nagercoil, 2015, pp. 1-4, doi: 10.1109/ICCPCT.2015.7159511.
(16) X. Liang et al., "Forest Data Collection Using Terrestrial Image-Based Point Clouds From a Handheld Camera Compared to Terrestrial and Personal Laser Scanning," in IEEE Transactions on Geoscience and Remote Sensing, vol. 53, no. 9, pp. 5117-5132, Sept. 2015. doi: 10.1109/TGRS.2015.2417316.
(17) B. Madore, G. Imahori, J. Kum, S. White and A. Worthem, "NOAA's use of remote sensing technology and the coastal mapping program," OCEANS 2018 MTSIEEE Charleston, Charleston, SC, 2018, pp. 1-7. doi: 10.1109/OCEANS.2018.8604932.
(18) Lyu, Yecheng, Lin Bai, Mahdi Elhousni, and Xinming Huang. "An Interactive LiDAR to Camera Calibration." arXiv preprint arXiv:1903.02122 (2019).
(19) Berrio, Julie Stephany, Mao Shan, Stewart Worrall, and Eduardo Nebot. "Camera-Lidar Integration: Probabilistic sensor fusion for semantic mapping." arXiv preprint arXiv:2007.05490 (2020).
(20) Domhof, Joris, and Kooij Julian FP. "An Extrinsic Calibration Tool for Radar, Camera and Lidar." In 2019 International Conference on Robotics and Automation (ICRA), pp. 8107-8113. IEEE, 2019.
(21) Sui, Jingfeng, and Shuo Wang. "Extrinsic calibration of camera and 3D laser sensor system." In 2017 36th Chinese Control Conference (CCC), pp. 6881-6886. IEEE, 2017.
(22) Zhou, Lipu, Zimo Li, and Michael Kaess. "Automatic extrinsic calibration of a camera and a 3d lidar using line and plane correspondences." In 2018 IEEE/RSJ International Conference on Intelligent Robots and Systems (IROS), pp. 5562-5569. IEEE, 2018.